\documentclass[%
twocolumn,
superscriptaddress,
showpacs,
nofootinbib,
nobibnotes,
amsmath,amssymb,
aps,
pra,
floatfix,
]{revtex4}
\usepackage{graphicx}
\usepackage{dcolumn}
\usepackage{bm}
\usepackage{hyperref}
\usepackage{natbib}
\usepackage{subfigure}

\begin{document}

\preprint{APS/123-QED}

\title{Two-Frequency Jahn-Teller Systems in Circuit QED}

%

\author{Tekin Dereli }
\affiliation{ Department of Physics, Ko\c{c} University, Sar{\i}yer,
Istanbul, 34450, Turkey}

\author{Yusuf G\"ul}
\affiliation{ Department of Physics, Ko\c{c} University, Sar{\i}yer,
Istanbul, 34450, Turkey}

\author{Pol Forn-D\'{i}az}
\affiliation{ Norman Bridge Laboratory of Physics, 
California Institute of Technology
Pasadena, CA 91125, USA}

\author{\"Ozg\"ur E. M\"ustecapl{\i}o\u{g}lu}
\email{omustecap@ku.edu.tr}
\affiliation{ Department of Physics, Ko\c{c} University, Sar{\i}yer,
Istanbul, 34450, Turkey}

%
\date{\today}

\begin{abstract}
We investigate the simulation of Jahn-Teller models with two non-degenerate vibrational modes using a circuit QED architecture.
Typical Jahn-Teller systems are anisotropic and require at least a two-frequency description. 
The proposed simulator consists of two superconducting lumped-element resonators interacting with a common
flux qubit in the ultrastrong coupling regime. We translate the circuit QED model of the system to a two-frequency Jahn-Teller Hamiltonian and calculate its energy eigenvalues and the emission spectrum of the cavities.
It is shown that the system can be systematically tuned to an
effective single mode Hamiltonian from the two-mode model by varying the coupling strength between the resonators.
The flexibility in manipulating the parameters of the circuit QED simulator permits isolating the effective single frequency 
and pure two-frequency effects in the spectral response of Jahn-Teller systems.  
\end{abstract}



\maketitle
\section{Introduction}

Simulating complex physical phenomena using systems that offer precise control of physical interactions, such as ultracold atoms \cite{lewenstein2006}, Bose-Einstein condensates \cite{stoof2001,garay2000} and trapped ions \cite{johanning2009} has attracted much attention in the last decade. It has been recently shown that cavity QED systems can be utilized for the same purpose, in particular to simulate certain gauge potentials, the anomalous Hall effect, and the Dirac equation \cite{larson2010}. The potential use of cavity QED systems to simulate such physical models relies on the successful simulation of  Jahn-Teller (JT) interactions \cite{yarkony1996,mead1992,bohm2003} which require atom-photon ultrastrong coupling conditions \cite{ciuti2006}. 

JT models describe the interaction of localized electronic states with vibrational (phonon) modes in crystals or in molecules \cite{bersuker2006}.
Cavity QED systems have been already proposed to simulate single mode JT models \cite{larson2008}.
On the other hand, many practical systems need a description in terms of multi-mode JT interactions \cite{markiewicz2002,majern2006a,majern2006,majernikova2011,eva2011}. We address the question of how to generalize  the restrictive single mode simulation of JT systems to two-mode JT interactions within the circuit QED context.

Circuit QED \cite{wallraff2004} offers the possibility to operate in the ultrastrong coupling regime \cite{pol2010,devoret2007,niemczyk2010,bourassa2009,daniel2011,milburn2010} 
for efficient JT and related spin-boson or Dicke model simulations. 
Photonic waveguide arrays are alternatively proposed \cite{crespi2011} for reaching the deep ultrastrong coupling regime (DSC) \cite{casanova2010} of JT type light-matter interaction.

Our idea is to consider a system consisting of a two-level atom simultaneously interacting with two cavities coupled to each other, rather than with a degenerate two-mode cavity, which was considered for the cylindrically symmetric  ${E}\times\epsilon$ JT model in cavity QED \cite{larson2008}. In terms of the normal modes of the coupled cavities, our system allows simulating a two-frequency (two non-degenerate vibrational normal modes) ${E}\times(\beta_1+\beta_2)$ JT model \cite{bersuker2006}. The normal modes of the two coupled cavities consist of a high frequency component and a low frequency component. The coupling strength between the cavities can be utilized to alter the frequency ratio of the modes to simulate different frequency ratios  encountered in different JT impurities in solids \cite{halperin1974}. In addition to more realistic simulations of JT systems, establishing a link between multi-mode JT models and coupled circuit QED systems could enable exploring many body physics such as quantum chaos \cite{majern2006a,majern2006}, quantum phase transitions \cite{majernikova2011}, and quantum entanglement in JT systems \cite{milburn2010, mckemmish2011} 
by using coupled cavity arrays.  

We consider a coupled circuit QED \cite{wallraff2004,chiorescu2004,reuther2010,matteo2008,matteo2011a,matteo2011b} system in the ultrastrong coupling regime as a feasible platform on which to realize our idea. The system consists of two coupled lumped-element LC resonators interacting with a two-level artificial atom, a superconducting flux qubit. In the ultrastrong coupling regime the rotating-wave approximation is not valid, so that the qubit-resonator coupling is of JT type 
rather than Jaynes-Cummings type \cite{bourassa2009,milburn2010},
allowing strongly coupled multi-frequency JT systems to be simulated.
The switchable ultrastrong coupling 
architecture can also be applied if a tunable coupling strength between the resonators and the flux qubit is desired \cite{peropadre2010}. 

Typical treatments of strongly coupled multi-mode JT systems in chemistry or in condensed matter physics utilize a cluster model \cite{vleck1939}, or use an effective single mode model where most of the JT interaction energy is concentrated predominantly upon a single  effective vibrational frequency with a negligible spread (narrow range of frequencies) \cite{obrien1972}. These methods are especially used for interpreting effects associated with low energy states, such as those seen in low temperature optical absorption \cite{fletcher1980}. 

When the frequency difference between the two modes is large, the situation is analogous to the case of optical and acoustic phonons in solids for which perturbative corrections become significant. 
We show that the frequency separation of the modes over which the JT interaction is distributed can be tuned with the coupling strength between the resonators. Our coupled circuit QED proposal allows for systematic simulation of effective single mode and pure two-mode effects as well as transitions between these regimes.  

This manuscript is organized as follows.  In Sec.~\ref{sec:model} we introduce the two-mode JT model and its implementation in a circuit QED context. The effective single mode treatment is described in Sec.~\ref{sec:effectiveModel}. In Sec.~\ref{sec:implem}, the experimental implementation is laid out. The results
and discussions are presented in Sec.~\ref{sec:results}. Finally we give conclusions in Sec.~\ref{sec:conclusion}.
\section{Two-Frequency JT Systems in cavity and circuit QED} \label{sec:model}
When two electronic levels are coupled to vibrations of ions or atoms in solids and molecules the general form of
the interaction can be written as $H_{JT}=\vec f(Q)\cdot\vec\sigma$, where $\vec f(Q)$ is a vector valued function of
vibrational coordinates while $\vec\sigma$ is the vector of the Pauli spin matrices. Such interactions are in general called
Jahn-Teller interactions \cite{bersuker2006}. In this paper, we focus on a particular one of the form $(Q_{1x}+Q_{2x})\sigma_x$, which is
known as the ${E}\times(\beta_1+\beta_2)$ JT (or Herzberg-Teller) model. 
Our aim is to generalize the recently discussed simulation of single mode
$E\times\epsilon$ JT model in cavity QED \cite{larson2008,larson2011}, which is of the form $Q_x\sigma_x+Q_y\sigma_y$, to the two-frequency case. Our
choice of $E\times(\beta_1+\beta_2)$ is the simplest possible two-frequency JT model.  
The implementation of $E\times(\beta_1+\beta_2)$ model allows for simulating realistic crystals that exhibit spatial
anisotropy. 
The single-boson $E\times\beta$ model is formally equivalent to the Dicke model and signatures of 
quantum chaos have been discussed in the $E\times(\beta_1+\beta_2)$ model recently \cite{majernikova2011}.

The Hamiltonian corresponding to the multi-frequency JT interaction between a single impurity ion and many vibrational 
degrees of freedom of the host lattice (or molecule) is expressed as
\begin{eqnarray} \label{eq:two-frequency JT model}
H=H_{ph}+H_{JT},
\end{eqnarray}
where $H_{ph}$ describes the free Hamiltonian of the phonon modes at frequencies $\omega_{i}$
\begin{eqnarray}
H_{ph}&=&\sum_{i}\hbar \omega_{i}(\hat a_{i}^{\dag} \hat a_{i}),
\end{eqnarray}
where $\hat a_{i} (\hat a_{i}^{\dag})$ are the annihilation (creation) operators of the phonons.
The multi-mode JT interaction describes the coupling of the single ion to the vibrational modes
\begin{eqnarray}
H_{JT}&=&\sum_{i}\hbar \omega_{i}k_i (\hat a_{i} + \hat a_{i}^{\dag})V.
\end{eqnarray}
Here $k_i$ are the dimensionless scaling factors of the JT coupling coefficients and $V$ is an operator that depends on the electronic degrees of freedom
of the impurity ion. 

We wish to simulate this multi-mode JT interaction using a coupled two-resonator circuit QED system. Normal modes of the coupled microwave photons  play the role of phonons, while a flux qubit plays the role of the impurity ion. The interaction of the flux qubit in the two-resonator circuit QED system mimics the local (short-range) interaction of the ion-phonon coupling. On the other hand, there is an additional non-local (long range) coupling between the resonator modes, describing hopping of photons between the resonators in the circuit QED system which mimics  the coupling between the vibrational phonons. The coupled resonator model can be written as ($\hbar=1$) 
\begin{eqnarray}\label{eq:two-resonator cirQED}
H=H_q+H_c+H_{qc}+H_{cc},
\end{eqnarray} 
where
\begin{eqnarray} 
H_q&=&\frac{\Omega}{2}\sigma_z,\\
H_{c} &=& \Omega_1\hat\alpha_1^\dag\hat\alpha_1+\Omega_2\hat\alpha_2^\dag\hat\alpha_2,\\
H_{qc} &=& [\lambda_1(\hat\alpha_1^\dag+\hat\alpha_1)+\lambda_2(\hat\alpha_2^\dag+\hat\alpha_2)]\sigma_x,\\
H_{cc} &=& J(\hat{\alpha_1}^{\dag}\hat{\alpha_2}+\hat{\alpha_2}^{\dag}\hat{\alpha_1}),
\end{eqnarray}
where $\sigma_z$, and $\sigma_x=\sigma_++\sigma_-$ are the Pauli spin operators describing the qubit degrees of freedom with  $\Omega$ being the qubit transition frequency, 
$\Omega_{1,2}$ the resonance frequencies of the cavity modes,  
$J$ the hopping rate of microwave photons between the resonators and $\hat\alpha_{1,2} (\hat\alpha_{1,2} ^\dag)$ the annihilation (creation) operators for the cavity photons. We want to simulate the two-frequency JT model in Eq.~\ref{eq:two-frequency JT model} with
the two-resonator circuit QED model in Eq.~\ref{eq:two-resonator cirQED}. For that aim it is necessary to be able to transform one model
to the other and show that they are identical for a certain set of model parameters. In the next section we examine the transformation between these two Hamiltonians. We apply the so-called effective single privileged mode transformation that has been developed for multi-frequency JT systems \cite{obrien1972} that allows for systematic analysis of pure single and multi-frequency effects.
\section{Effective Single-mode JT system in two-resonator circuit QED} \label{sec:effectiveModel}

We now employ the effective single mode treatment \cite{obrien1972} for the two-frequency JT model obtained within the two-resonator circuit QED context.  For that aim we look for a particular linear superposition the of normal modes of the coupled resonators,
\begin{equation} \label{eq:transformation}
\hat \alpha_{i}=\sum_{k}A_{ik}\hat a_{k},
\end{equation}
where $A_{ik}$ are the elements of a real orthogonal matrix to be determined, such that most of the JT energy is concentrated over a privileged mode among the set of new bosonic modes $\hat\alpha_{i}$.

Without loss of generality, we choose the privileged mode as $\hat\alpha_{1}$ for which the single-mode model JT system can be written in terms of an effective frequency $\omega_{eff}$ and an effective JT coupling $k_{eff}$, scaled by 
$\omega_{eff}$, so that
\begin{eqnarray}\label{eq11}
H_{eff}&=&\frac{\Omega}{2}\sigma_{z}+\omega_{eff}[\hat \alpha_{1}^{\dag}\hat \alpha_{1}+k_{eff}(\hat \alpha_{1}+\hat \alpha_{1}^{\dag})V].
\end{eqnarray}

The relevant elements of the transformation, $A_{11}$ and $A_{12}$, are determined by maximizing $k_{eff}^2\omega_{eff}$. This is the amount by which the minimum of the potential energy of the system is lowered, under adiabatic approximation, due to the interaction of the rest of the system with such a single-mode \cite{obrien1972}. This
gives $A_{11}:A_{12}=k_1:k_2$, subject to normalization conditions. Direct substitution of 
Eq.~\ref{eq:transformation} into Eq.~\ref{eq:two-frequency JT model} yields
\begin{eqnarray}
\omega_{eff}&=&\frac{\omega_1k_1^2+\omega_2k_2^2}{k_{eff}},\\
k_{eff}^2&=&k_1^2+k_2^2.
\end{eqnarray}

For a $2\times 2$ orthogonal matrix $A$ determining the first row of elements fixes the remaining two by orthonormality conditions such that
\begin{equation} A=\frac{1}{k_{eff}}
\left(
                        \begin {array}{ccc}
                        k_{1} & k_{2} \\
                        k_{2} &  -k_{1}
                        \end{array}
\right),
\end{equation}
which is taken in consistency with Ref.~\cite{obrien1972}. 
Such transformations are
common in Morris-Shore bright and dark state transformations \cite{morris1983}. For the system analyzed here, on the contrary, there is no perfect decoupling of either mode from the dynamics of the rest of the system,
though for certain parameter regimes the modes $\hat\alpha_1$ and $\hat\alpha_2$ become approximately decoupled.
The total transformed Hamiltonian can be written as
\begin{eqnarray}
\tilde{H}_{sys}&=&H_{eff}+H'_{ph}+H_{int},
\end{eqnarray}
where
\begin{eqnarray}
H'_{ph}&=&\omega'\hat \alpha_{2}^{\dag}\hat \alpha_{2},
\end{eqnarray} with
\begin{eqnarray}
\omega'=\frac{\omega_1k_2^2+\omega_2k_1^2}{k_{eff}},\\
\end{eqnarray}
 is the free Hamiltonian of the disadvantaged effective mode. Interaction of this mode with the rest of the system is described by
\begin{eqnarray}\label{eq19}
H_{int}=c_{2}[(\hat \alpha_{1}^{\dag}\hat \alpha_{2}+\hat \alpha_{1}\hat \alpha_{2}^{\dag})
+k_{eff}(\hat \alpha_{2}+\hat \alpha_{2}^{\dag})V].
\end{eqnarray}
Here the strength of the coupling between the privileged and the disadvantaged modes is characterized by the parameter
\begin{eqnarray}
c_{2}=\Delta\frac{k_1k_2}{k_{eff}^2},
\end{eqnarray}
where $\Delta=\omega_1-\omega_2$ is the frequency difference between the vibration modes in the two-frequency JT model.
If the vibrational modes are degenerate the model is exactly equivalent to the case of an effective single frequency. 
The coupling between the JT vibration modes can also be expressed as
$c_{2}^{2}=\overline{\omega^{2}}-\overline{\omega}^{2}$,
where
\begin{eqnarray}
\overline{\omega^{n}}=\frac{\omega_{1}^{n}k_{1}^{2}+\omega_{2}^{n}k_{2}^{2}}{k_{1}^{2}+k_{2}^{2}},
\end{eqnarray}
This allows for interpreting $c_2$ as the mean square width of that distribution with a mean $\overline{\omega}=\omega_{eff}$ \cite{obrien1972}.

The effect of $H_{int}$ on the effective single-mode model can be examined perturbatively provided that the frequency spread ($\Delta$ or $c_2$)  is not too large. Perturbative effects will only be significant on the JT ground state only starting from the second order. To see this it is convenient to introduce a new set of operators \cite{obrien1972}
\begin{eqnarray}
\hat\eta=\hat\alpha_{1}+k_{eff}V,
\end{eqnarray}
for which we can re-express the effective model as
\begin{eqnarray}
H_{eff}=\omega_{eff}\hat\eta^\dag\hat\eta-\omega_{eff}k_{eff}^2V^2.
\end{eqnarray}
The last term is proportional to the unit matrix. $\hat\eta$ obeys bosonic commutation rules even though it contains
Pauli spin operators. 
This operator can be identified as a ``bright" qubit-polariton quasi-particle. The first term in the effective Hamiltonain is then a 
harmonic oscillator contribution due to the free energy of these quasi-particles.
The interaction Hamiltonian becomes
\begin{eqnarray}
H_{int}=c_{2}(\hat\eta \hat\alpha_{2}^{\dag}+\hat\eta^{\dag}\hat\alpha_{2}),
\end{eqnarray}
representing the coupling of a bright qubit-polariton to a ``dark" effective mode $\hat\alpha_2$.
Using commutation relations between $H_{eff}$ and $\hat\alpha_1$, the necessary matrix elements of $\hat\eta$ for the 
perturbation analysis in the representation of eigenstates of $H_{eff}$ can be determined \cite{obrien1972}. The simple relation
$[H_{eff},\hat\alpha_1]=-\omega_{eff}\hat\eta$ shows that $\hat\eta$ has no diagonal elements. This means that the perturbative corrections can only be of significance starting from second order, at least for the JT ground state and for low energy lying states. If the frequency spread $c_2$ is not negligible compared to the Jahn-Teller coupling $k_{eff}$ the description of the system using a privileged single mode is not possible since the JT energy is spread among the two modes $\hat\alpha_1,\hat\alpha_2$.

We are now in a position to relate the two-resonator circuit QED model of Eq.~\ref{eq:two-resonator cirQED} to the two-frequency JT
model of Eq.~\ref{eq:two-frequency JT model}. We find that the parameters of the Hamiltonians are related as follows
\begin{eqnarray}\label{eq:relations}
\omega_{eff}&=&\Omega_1, \quad \omega^\prime=\Omega_2\quad \omega_{eff}k_{eff}=\lambda_1,\nonumber\\
c_2k_{eff}&=&\lambda_2,\quad c_2=J.
\end{eqnarray}
The relations require that a condition of the form 
\begin{eqnarray}\label{eq:condition}
\Omega_1=\frac{\lambda_1}{\lambda_2}J
\end{eqnarray}
should be satisfied among the parameters of the circuit QED Hamiltonian.
\section{Experimental implementation}\label{sec:implem}
\begin{figure}[!hbt]
\begin{center}
\includegraphics[width=0.4\textwidth]{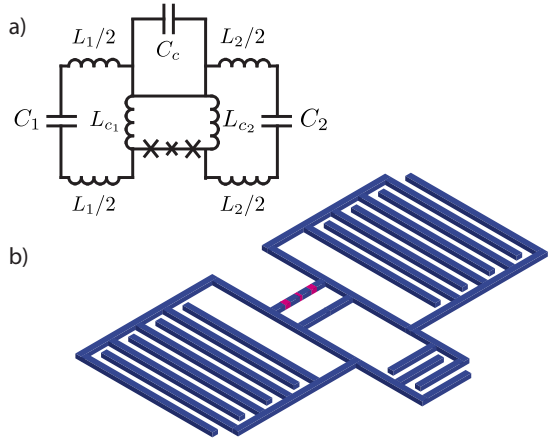}
\caption{\label{fig:circ} a) (Color online) Schematic of a flux qubit galvanically coupled to two $LC$ resonators through coupling inductances $L_{c_{1,2}}$. Each resonator has resonant frequency $\omega_{1,2}=((L_{1,2}+L_{c_{1,2}})C_{1,2})^{-1/2}$. The resonators are coupled to each other through the coupling capacitor $C_c$. b) Implementation of the circuit using interdigitate finger capacitors. Feed lines individually coupled to each resonator can be employed to probe the internal photon state of the resonators.}
\end{center}
\end{figure}

The system we consider to implement the Hamiltonian in Eq.~\ref{eq:two-resonator cirQED} consists of two lumped-element $LC$ resonators capacitively coupled to each other and a flux qubit simultaneously coupled to each resonator. A schematic of the circuit can be seen in Fig.~\ref{fig:circ}. 
The resonator-resonator interaction $J$ is determined by the coupling capacitor $C_c$ between the two $LC$ resonators when they are in the ground state \mbox{$J\simeq C_{c}V_{\mathrm{rms}_1}V_{\mathrm{rms}_2}=C_{c}\sqrt{\frac{\omega_1\omega_2}{4C_1C_2}}$}, where $C_{1,2}$ is the total capacitance of the uncoupled resonators. Using typical sample parameters \cite{pol2010, chiorescu2004}, the coupling strength between the two resonators can be made very large, up to a considerable fraction of the frequency of each resonator.
The spurious inductive resonator-resonator coupling could be minimized with an appropriate resonator geometry, if necessary. The flux qubit can also be made of large enough size (as the ones in \cite{hime2006}) so as to increase the distance between the resonators and reduce the mutual resonator-resonator inductance.

The coupling energy $\lambda$ in a flux qubit-resonator system can become a large fraction of the energy of the resonator if the qubit is galvanically attached to the resonator, as already demonstrated experimentally in Refs.~\cite{niemczyk2010, pol2010}. For a qubit either sharing a long section of its inductance \cite{pol2010} or a Josephson junction \cite{niemczyk2010, bourassa2009} with a resonator, coupling energies $\lambda\approx \omega_{1,2}$ are within reach experimentally.

From the analysis of the sec.~\ref{sec:effectiveModel}, in order to study the privileged mode regime the coupling term in Eq.~\ref{eq11} needs to be larger than the coupling term in Eq.~\ref{eq19}. This implies, according to the relations in Eq.~\ref{eq:relations}, that $\lambda_1>\lambda_2,J$. Therefore in the experiment the flux qubit has to be ultrastrongly coupled to one resonator and strongly coupled to the other resonator, while the resonator-resonator coupling must be close to the qubit-resonator strong coupling. These designed coupling energies will determine the privileged mode. By detecting the photon state of each resonator using feed lines \cite{fink2010} permits exploring the spectral properties of the complete system.



\section{Results} \label{sec:results}
For the sake of simplicity, we choose $k_1=k_2=k$ so that $k_{eff}=\sqrt{2}k$. The relations in Eq.~\ref{eq:relations} reduce to
$\Omega_1=\Omega_2\equiv\Omega_c=(\omega_1+\omega_2)/2, \lambda_1=(\omega_1+\omega_2)k/\sqrt{2},
\lambda_2=\Delta k/\sqrt{2}$, and $J=c_2=\Delta/2$. Our choice requires the resonators in the circuit QED system 
to be degenerate. We further assume resonance condition $\Omega=\Omega_c$. The circuit 
QED Hamiltonian then becomes
\begin{eqnarray} \label{eq:cirQED model}
H&=&\hat\alpha_1^\dag\hat\alpha_1+\hat\alpha_2^\dag\hat\alpha_2+\frac{1}{2}\sigma_z+\frac{\Delta}{2}(\hat\alpha_1^\dag\hat\alpha_2
+\hat\alpha_2^\dag\hat\alpha_1)\nonumber\\
&+&k_{eff}[(\hat\alpha_1^\dag+\hat\alpha_1)+\frac{\Delta}{2}(\hat\alpha_2^\dag+\hat\alpha_2)]\sigma_x.
\end{eqnarray}
We use dimensionless energy and time, respectively scaled by $\hbar\Omega_c$ and $1/\Omega_c$, but do not change our notation for scaled variables.  Our model is then a two-parameter ($k,\Delta$) theory where $\Delta$ is in units of $\Omega_c$. 

The resonators are degenerate but the system still simulates the two-frequency JT model. The coupling coefficient between the resonators $J$ determines the frequency ratio of the two vibration modes in the corresponding two-frequency JT Hamiltonian 
of Eq. \ref{eq:two-frequency JT model} to be simulated, which becomes
\begin{eqnarray}
H&=&\omega_1\hat a_1^\dag\hat a_1+\omega_2\hat a_2^\dag\hat a_2+\frac{1}{2}\sigma_z+\nonumber\\
&+&k[\omega_1(\hat a_1^\dag+\hat a_1)+\omega_2(\hat a_2^\dag+\hat a_2)]\sigma_x,
\end{eqnarray}
where the frequency ratio is determined by
\begin{eqnarray}
\frac{\omega_1}{\omega_2}=\frac{1+\Delta/2}{1-\Delta/2}.
\end{eqnarray}
Some cases of interest are the 2:1 frequency ratio of the two phonon modes in C$_6$H$_6$$^{\pm}$, and 
the frequency ratio $3:1$ of the two phonon modes of Fe$^{2+}$ in ZnS \cite{halperin1974}. As explained in sec.~\ref{sec:effectiveModel}, the transformation or equivalence of these two Hamiltonians is exact. 
For $\Delta$ (or the frequency spread $c_2$) small compared to the uncoupled eigenfrequencies of the qubit and the resonators, we obtain a faithful representation of the two-frequency JT model in terms of the privileged mode.
Corrections appear only as second order perturbations. 

\begin{figure}[!hbt]
\begin{center}
\subfigure[\hspace{0.01cm}]{\label{fig:2a}
\includegraphics[width=0.4\textwidth]{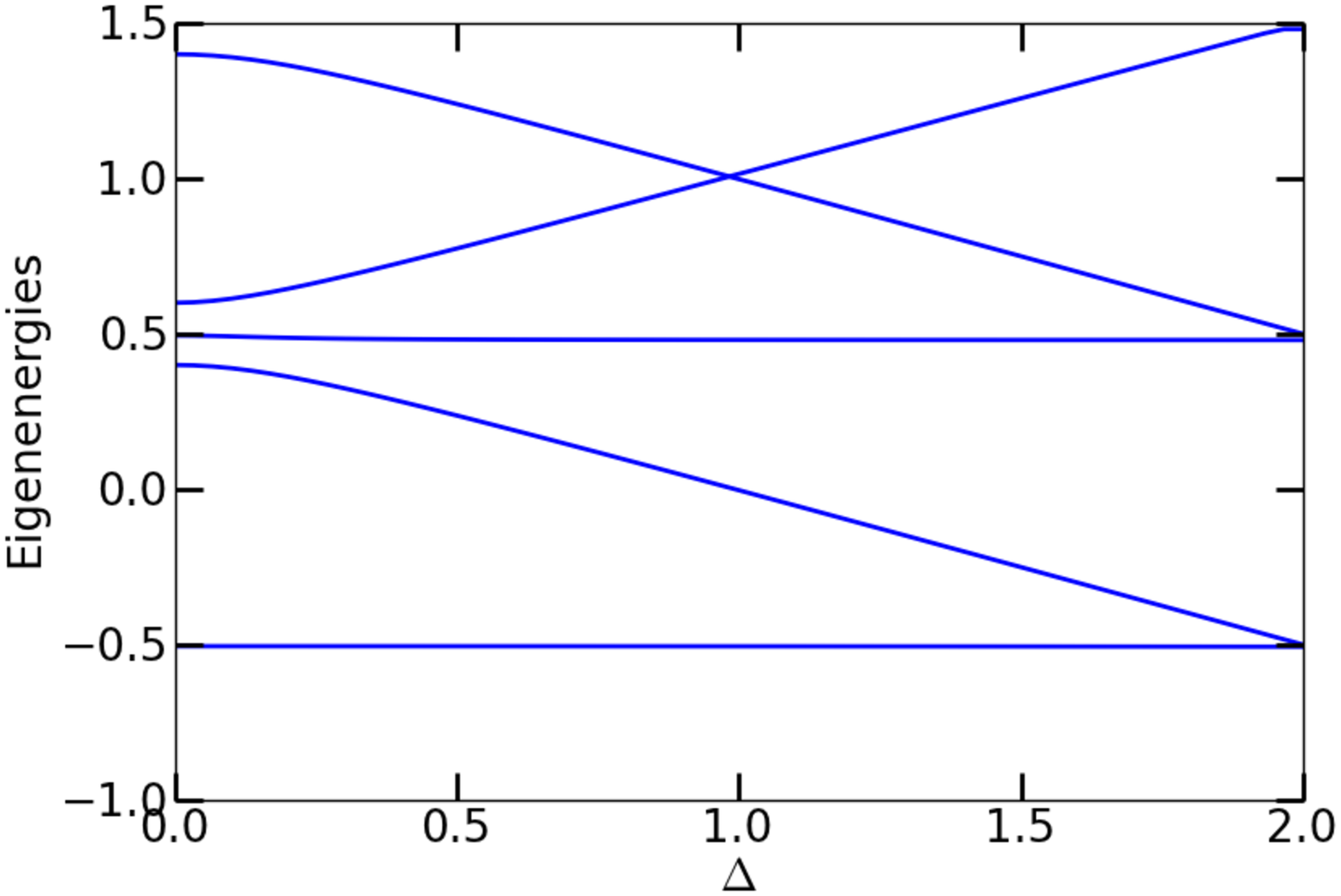}}
\subfigure[\hspace{0.01cm}]{\label{fig:2b}
\includegraphics[width=0.4\textwidth]{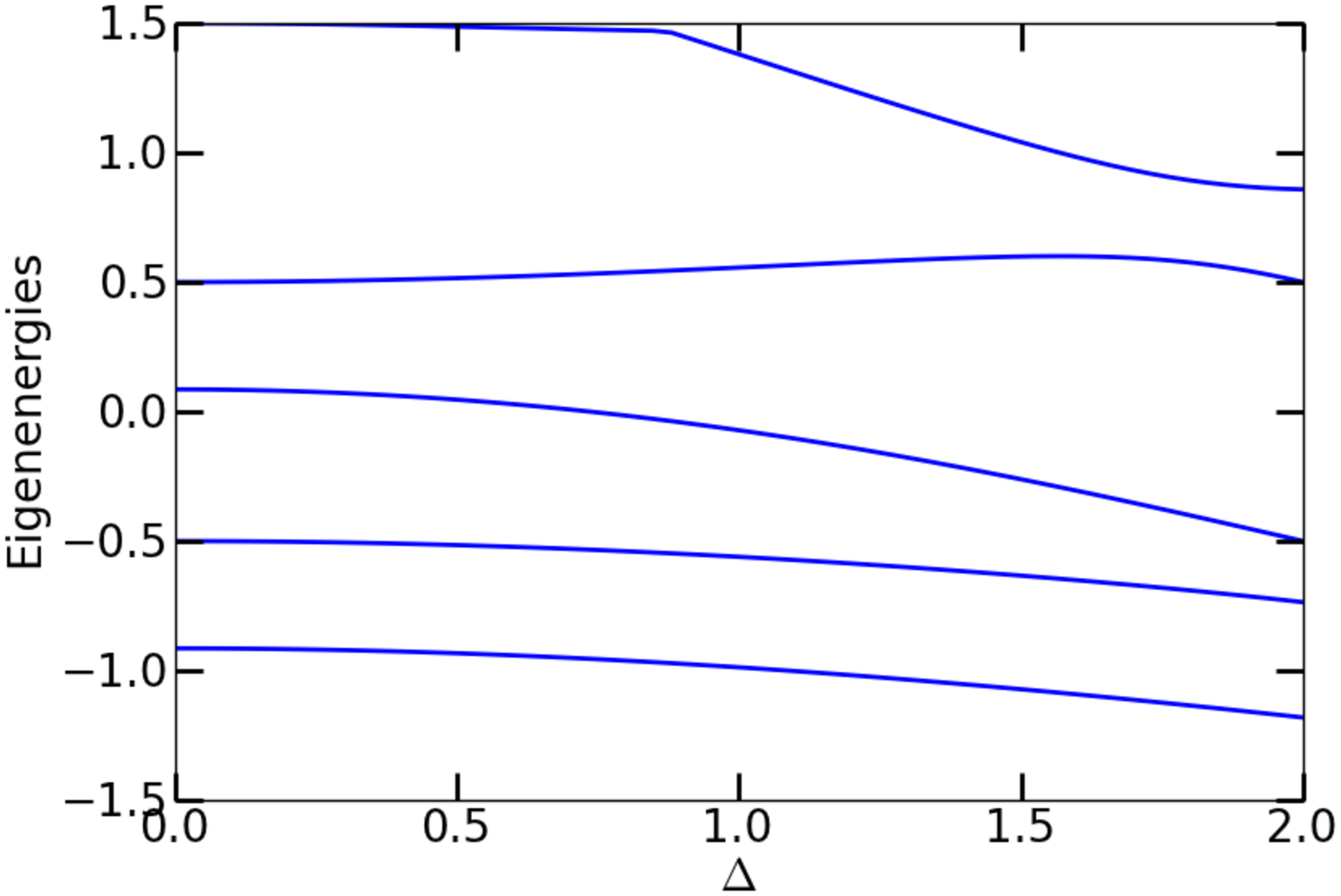}}
\subfigure[\hspace{0.01cm}]{\label{fig:2c}
\includegraphics[width=0.4\textwidth]{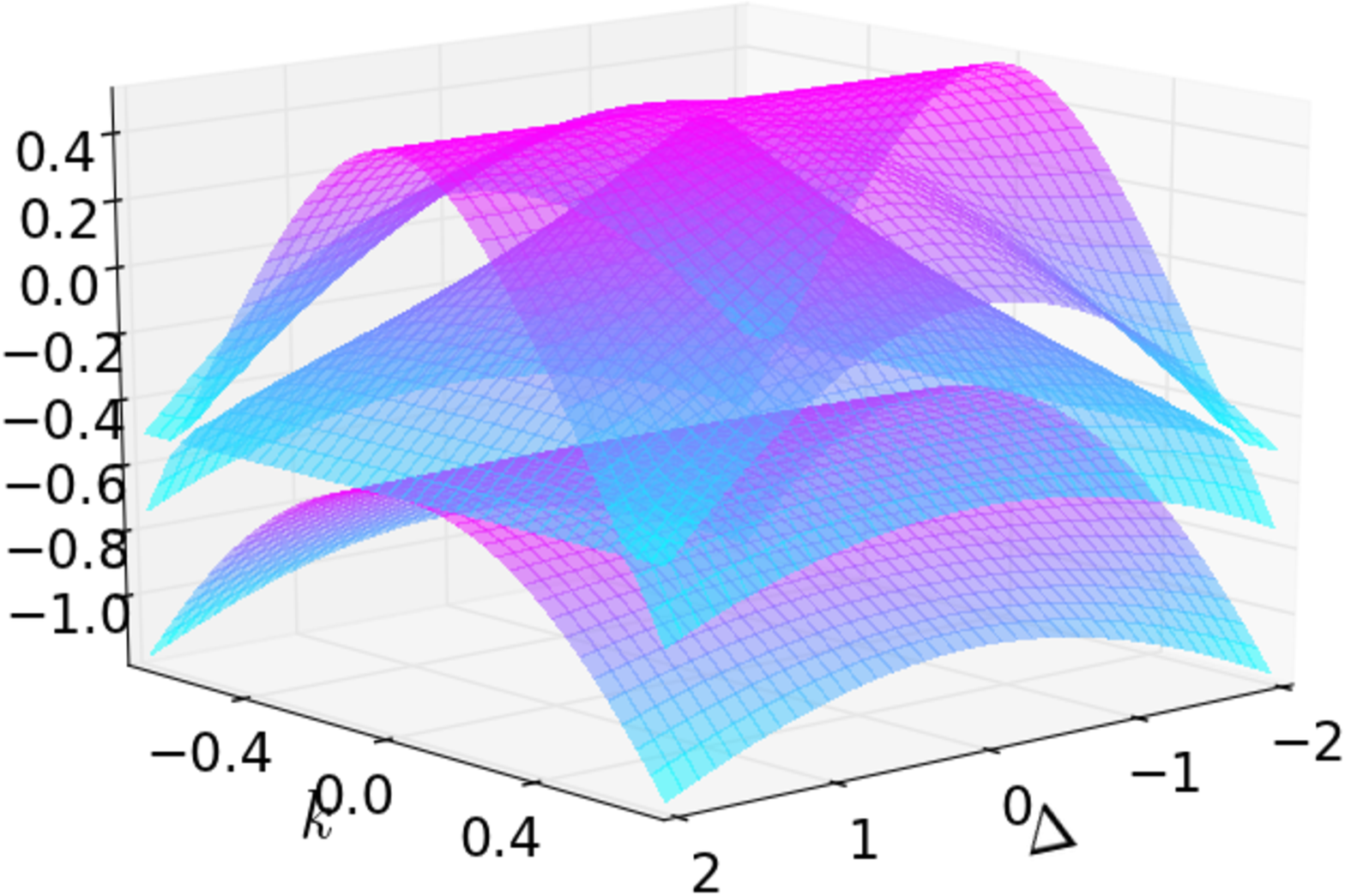}}
\caption{\label{fig2} (Color online)  Dependence of the lowest five eigenenergies of the two-frequency Jahn-Teller model on the 
mode coupling parameter $\Delta$. (a) For the
qubit-cavity coupling scaling parameter $k=0.1/\sqrt{2}$, the splitting of the first energy level into three at $\Delta=0$ is a pure Rabi splitting. The privileged effective single mode regime is valid over a narrow band $|\Delta| < 0.1$. (b) At $k=1/\sqrt{2}$ the energy-level transitions become flat over a broader band $|\Delta| < 1$, deeper into the ultrastrong coupling regime. (c) Energy bands of the two-frequency 
Jahn-Teller model depending on the mode coupling
parameter $\Delta$ and the qubit-cavity coupling scaling parameter $k$.  
All the quantities plotted are dimensionless as explained in the text.}
\end{center}
\end{figure}

We first examine the eigenenergies of the Hamiltonian in Eq.~\ref{eq:cirQED model}. The lowest five eigenvalues are shown in
Fig.~\ref{fig2}. 
Here the Fock space dimensions for each resonator mode is fixed to $2$ so that we consider up to two-photon
manifolds. We examined the influence of dimensions of the Fock space on our results in the case of spectrum calculations and found them to be sufficiently robust. The strong coupling case with $k=0.1/\sqrt{2}$ (or just the beginning of the ultrastrong coupling regime) is considered in
Fig.~\ref{fig:2a}. When $\Delta=0$ there is only pure Rabi splitting as can be seen in the first excited level. When $\Delta$ increases,
the coupling between the privileged and the disadvantaged modes increases. This polaritonic interaction of the modes causes further repelling of the Rabi-split levels. Single privileged effective mode description of the system would only be valid over a narrow band $k\sim|\Delta|<0.1$. The ultrastrong coupling regime with $k=1/\sqrt{2}$
is considered in Fig.~\ref{fig:2b}. An asymmetric Rabi splitting at $\Delta=0$ can be seen in the first excited level. Here the effective single
mode description is valid over a broader range $|\Delta|<1$. The dependence of the energy spectrum on the 
full range of $k$ and $\Delta$ is shown in Fig.~\ref{fig:2c}. The first band is tent-shaped, and for low $k$ it varies sharply with $\Delta$
resulting in a narrow regime of the effective single mode description. As $k$ reaches ultrastrong coupling conditions, the regime of effective
single mode description becomes more robust against variations in $\Delta$ over a broader range.  

Solid state and molecular multi-frequency JT systems are usually investigated through their absorption spectrum. The corresponding quantity 
in circuit QED is the transmission spectrum of the resonators. We consider the power spectrum of only one resonator, corresponding to the privileged mode. Deviations from single mode behavior in this spectrum would be identified as pure two-frequency effects. In order to calculate the power spectrum it is necessary to solve the quantum master equation for the ultrastrong coupling regime, which can only be rigorously formulated in the dressed state picture of coupled qubit-resonator, examined recently in refs.~\cite{beaudoin2011,hausinger2008}. Our purpose is to see qualititative changes in the spectrum at different frequency ratios of the two-frequency JT model. We assume that the usual Bloch-Redfield quantum master equation for circuit QED systems in the Born-Markov approximation is applicable for our purposes \cite{fink2010}, and the equation is given as ($\hbar=1$)
\begin{eqnarray}
\frac{d\rho}{dt}=-i[H,\rho]+{\cal L}\rho,
\end{eqnarray}
where the Liouvillian superoperator ${\cal L}$ is given by
\begin{eqnarray}
 {\cal L}\rho&=&\sum_{j=1,2}(1+n_{th})\kappa{\cal D}[\hat\alpha_j]\rho+n_{th}\kappa{\cal D}[\hat\alpha_j^\dag]\rho\nonumber\\
 &+&\gamma{\cal D}[\sigma]\rho+\frac{\gamma_\phi}{2}{\cal D}[\sigma_z]\rho,
 \end{eqnarray}
 with $n_{th}$ being the average thermal photon number, which we take as $n_{th}=0.1$ corresponding to $100$~mK \cite{fink2010}. 
 The Lindblad type damping superoperators are denoted by ${\cal D}$. The cavity photon loss rate $\kappa$ is taken to be the same
 for both resonators. The qubit relaxation and dephasing rates are represented by $\gamma$ and $\gamma_\phi$, respectively. 
  
The JT spectrum is determined by calculating the real part of the Fourier transform
of the stationary two-time first-order correlation function for the privileged mode $\hat\alpha_1$, so that
\begin{eqnarray}
P(\omega)=\int_{-\infty}^{\infty}\langle\hat\alpha_1^\dag(t)\hat\alpha_1(0)\rangle e^{-i\omega t}.
\end{eqnarray}
We use Python programming language with the QuTip package for the determination of the spectrum 
\cite{qutip}. The decay parameters, scaled by $\Omega_c$, are taken to be $\kappa=0.001,\gamma=0.001$ and $\gamma_\phi=0.01$, while assuming Fock space of up to two
photons for each resonator mode. When we consider higher Fock space dimensions (due to numerical constraints we examined up to 5 particle manifolds) we find that the spectrum is robust against the variations in the Fock space dimensions for the regimes of $J\le 0.5$ we consider. At larger $J$ values small changes in the spectral intensities are observed, but they are still negligible up to $J\sim 1$. For smaller decay rates and at even larger $J$ values the spectrum becomes more sensitive to dimensions of the Fock space.
\begin{figure}[!]
\begin{center}
\subfigure[\hspace{0.01cm}]{\label{fig:3a}
\includegraphics[width=0.4\textwidth]{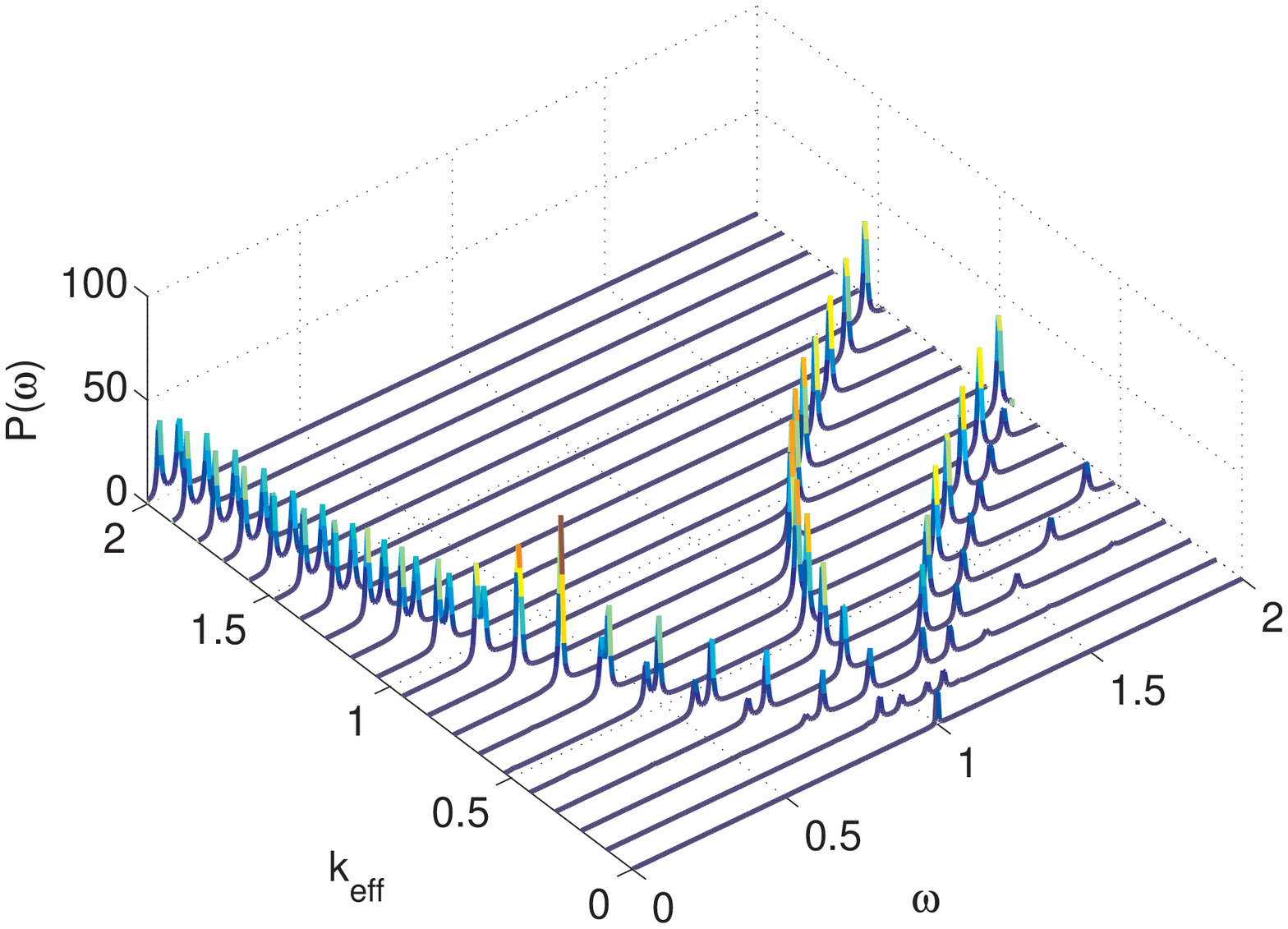}}
\subfigure[\hspace{0.01cm}]{\label{fig:3b}
\includegraphics[width=0.4\textwidth]{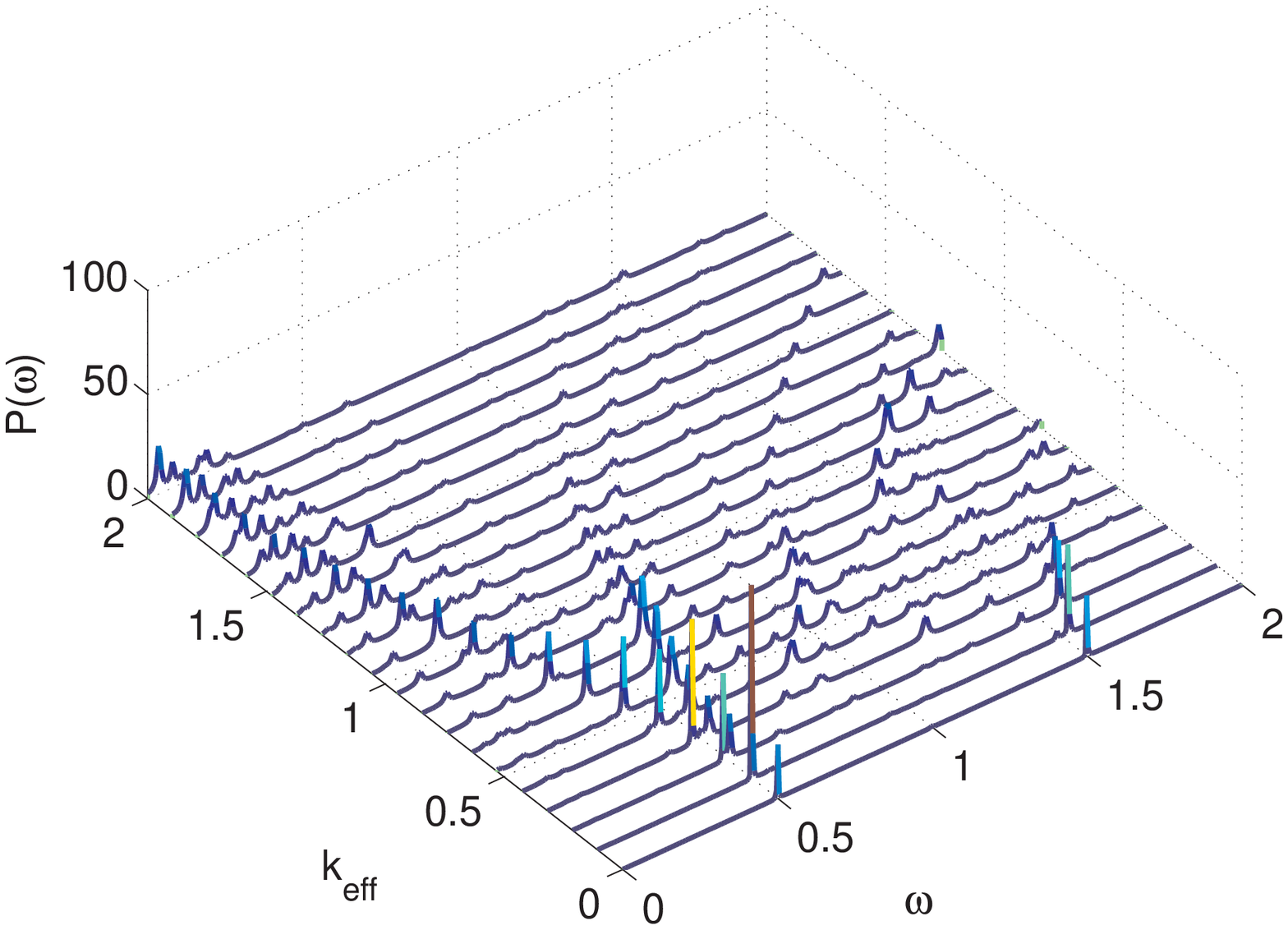}}
\subfigure[\hspace{0.01cm}]{\label{fig:3c}
\includegraphics[width=0.4\textwidth]{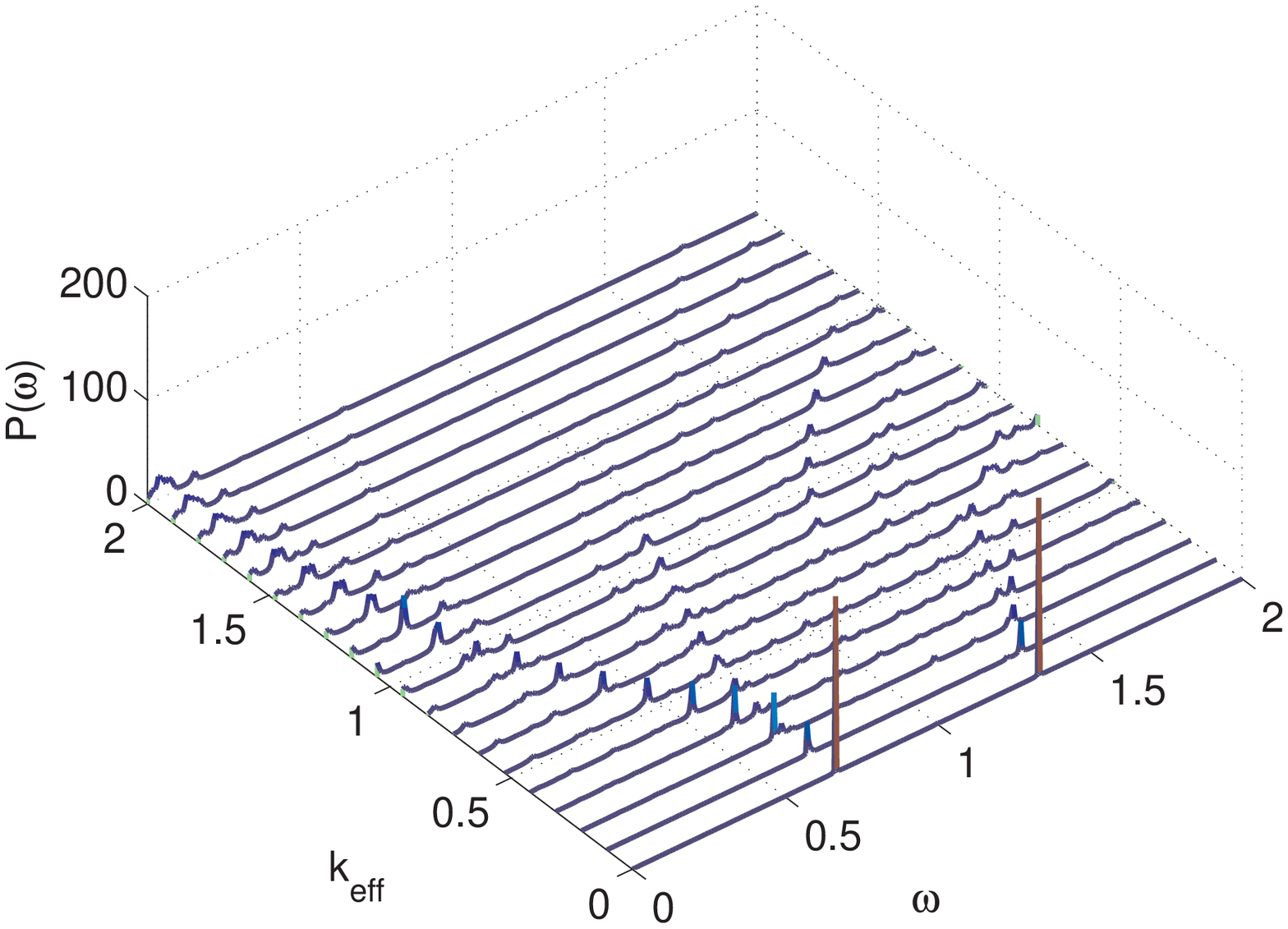}}
\caption{\label{fig3} (Color online)  Resonator-qubit coupling $k_{eff}$ dependence of the cavity emission spectrum 
for the privileged effective 
mode in the two-frequency Jahn-Teller model for different values of the resonator-resonator coupling strength $J$. (a) $J=0$ corresponds to the pure single-mode regime. $k_{eff}\ll 1$ is the single-mode regime in the JC model. For $k_{eff}\gtrsim 1$ one enters the regime of single-mode JT model. (b) $J=1/2$ shows three regimes: $k_{eff}\ll 1$ (2-mode JC model), $k_{eff}\gtrsim 1$ (2-mode JT model), and $k_{eff}\gg J$ (effective single-mode JT model). This case simulates the frequency ratio of $3:1$ of the two phonon modes in Fe$^{2+}$ in ZnS. (c) $J=1/3$ simulates the frequency ratio of $2:1$ of the two phonon modes in C$_6$H$_6$$^{\pm}$. In this case the effective single-mode regime is more clear for $k_{eff}>1.5$.}
\end{center}
\end{figure}

Our results for different values of $J$ are presented in Fig.~\ref{fig3}. Fig.~\ref{fig:3a} shows the spectrum when the two resonators are uncoupled, $J=0$.
For low $k_{eff}$, the spectrum shows typical asymmetric Rabi-split frequency peaks of the Jaynes-Cummings (JC) model around the degenerate frequency of the resonators. At larger values of $k_{eff}\sim 1$ the system is in the single-mode JT regime. Fig.~\ref{fig:3b} presents the effect of coupling the resonators with $J=1/2$, which corresponds to the typical frequency ratio $3:1$ of two phonon modes of Fe$^{2+}$ in ZnS.
The polaritonic splitting here shifts the Rabi-split peaks further away at low values of $k_{eff}$. For $k_{eff}\sim 1$ the system is in the two-mode JT regime. Fig.~\ref{fig:3c} shows similar features for $J=1/3$, which corresponds to the typical value of frequency ratio $2:1$ of the two phonon modes in C$_6$H$_6$$^{\pm}$.
In this case for $k_{eff}>1.5$ the system enters the effective single mode JT regime. 

\begin{figure}[!]
\begin{center}
\subfigure[\hspace{0.01cm}]{\label{fig:4a}
\includegraphics[width=0.4\textwidth]{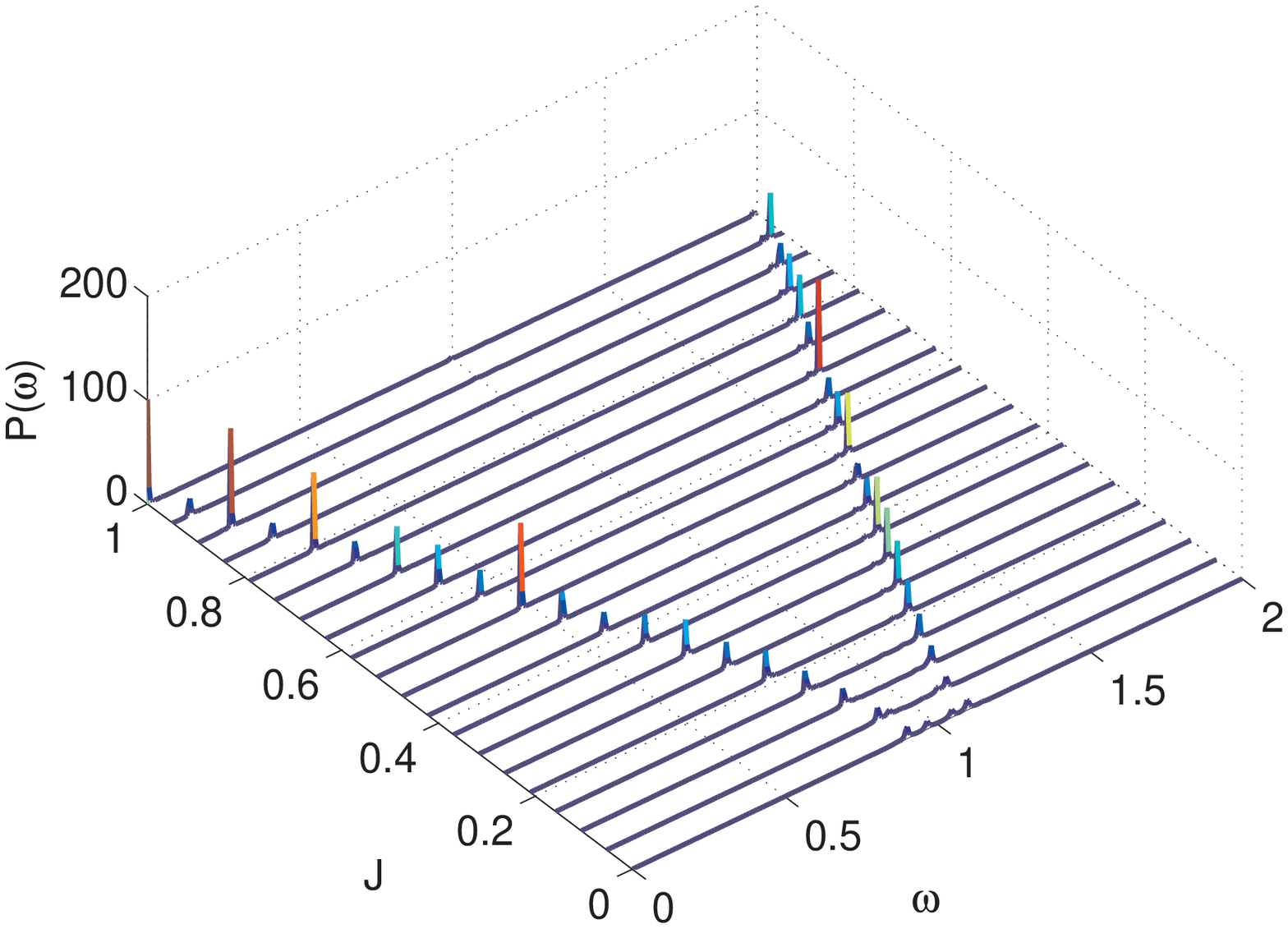}}
\subfigure[\hspace{0.01cm}]{\label{fig:4b}
\includegraphics[width=0.4\textwidth]{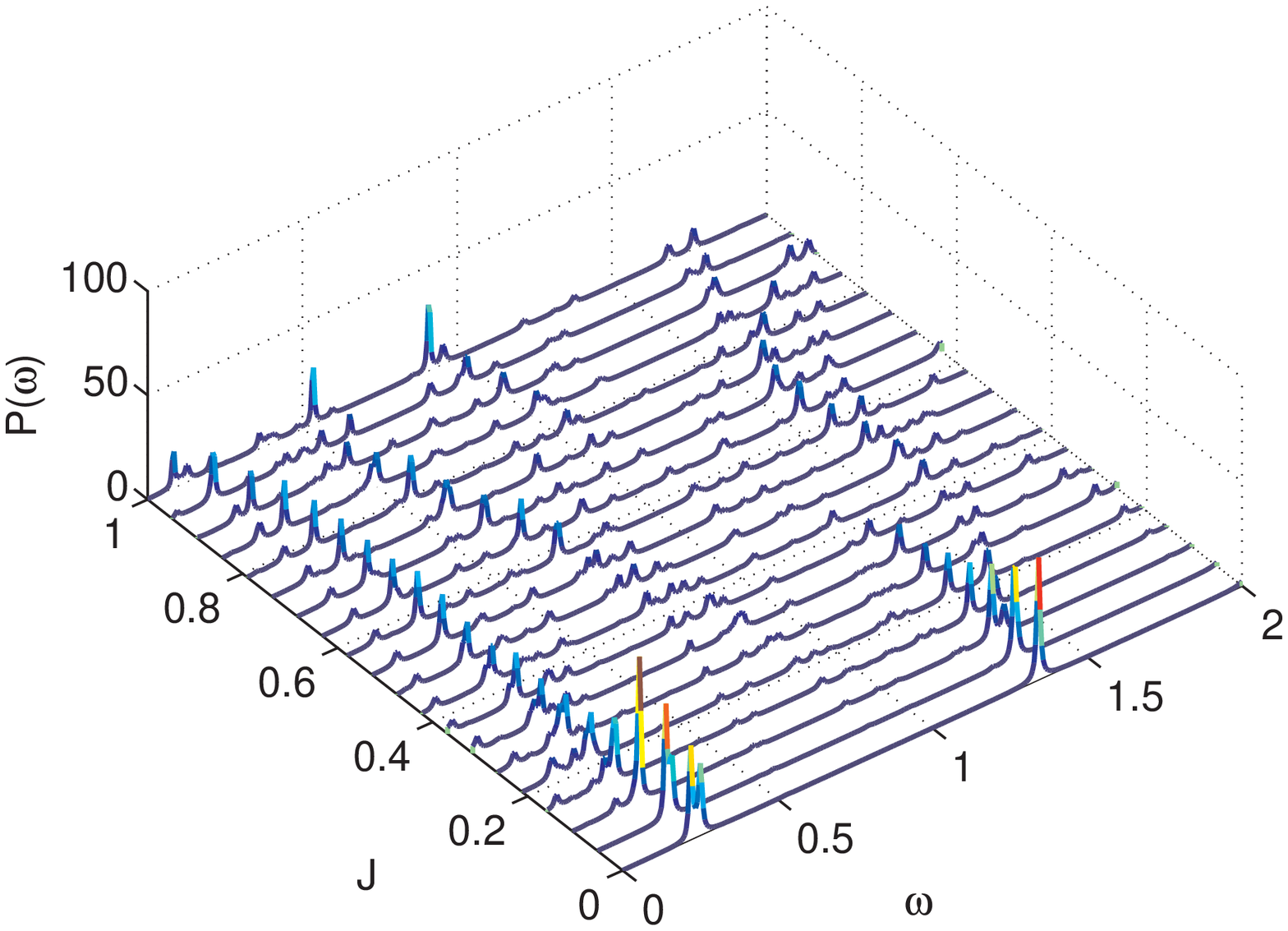}}
\caption{\label{fig4} (Color online)  Variation of cavity emission spectrum for the privileged effective 
mode in the two-frequency Jahn-Teller model with resonator-resonator coupling 
$J$ at (a) $k_{eff}=0.1$ and (b) $k_{eff}=1$.}
\end{center}
\end{figure}

The general behavior of the spectrum with the resonator-resonator coupling $J$ at a given qubit-resonator coupling $k_{eff}$ is shown in Fig.~\ref{fig4}. 
Fig.~\ref{fig:4a} shows the case when $k$ is near the threshold of the ultrastrong coupling regime, $k_{eff}=0.1$. The spectrum is mainly determined by the two-frequency character of the system, corresponding to the two-mode Jaynes-Cummings model. The privileged single mode description is limited to very small coupling strengths $J=\Delta/2\lesssim0.1$. Beyond this point, normal mode splitting increases with $J$. 

When we consider the case of deeper ultrastrong coupling $k_{eff}=1$, Fig.~\ref{fig:4b} reveals that it is easier to resolve the single-mode/two-mode regimes as the single-mode regime 
is significantly enhanced, up to $J=\Delta/2\sim 0.4<k_{eff}$. This observation complies with our previous arguments, 
based upon the energy levels of the system in 
Fig.~\ref{fig3}. Beyond the single mode regime, Fig.~\ref{fig:4b} shows that 
the higher frequency peak at $\omega\sim 1.4$ disappears, while the lower frequency one at $\omega\sim 0.2$ dominates. 
The spectrum exhibits additional peaks that grow in number and in amplitude in the two-frequency regime. 
These peaks are due to multi-photon processes that become more and more significant as one goes deeper into 
the strongly coupled JT model \cite{naderi2011}.
The higher energy resonance in the single mode regime turns out to be more susceptible to such multi-photon processes. 
The amplitude of this transition decreases and eventually vanishes in the two-frequency regime, while the lower energy resonance is more robust
and does not decrease its amplitude significantly. 
These results suggest that one can monitor and analyze the 
transition between the effective single privileged mode and  the pure two-frequency behavior of a JT system
by tuning the circuit QED parameters into the DSC regime. 
\section{CONCLUSION}\label{sec:conclusion}
In summary, we have presented a method to simulate a two-frequency JT model by using a two-resonator circuit QED system. The proposed model consists of a flux qubit coupled to two resonators in the ultrastrong coupling regime. An exact transformation between the two-frequency JT Hamiltonian and the circuit QED Hamiltonian has been established. The transformation permits describing the system in terms of an effective privileged single mode under certain conditions of the control parameters of the circuit QED system. The effective disadvantaged mode can be de-coupled from the privileged one in the ultrastrong coupling regime. 
The eigenenergy spectrum and power spectrum are calculated using ultrastrong circuit QED parameters, with specific
 attention to the present experimental restrictions. The tunability of the pure two-mode JT model and the effective privileged mode model is found to be feasible in the ultrastrong coupling circuit QED within the range of parameters in present experiments. 
 
Simulating and interpreting more complex JT systems, such as vacancies in graphite or fullerides  C$_{60}^{-}$, would require going beyond a two-mode description. Our analysis of the two-frequency JT model simulation can serve as a building block for further realizations of other classes of multi-mode JT systems, by considering for example 
 coupled multi-mode superconducting transmission line resonators and their interactions with flux qubits in the (ultra)strong coupling regime. Such extensions of the present work would allow examining rich geometric phase effects \cite{larson2011} and designing synthetic gauge fields, as well as enhancing the comprehension of nonlinear JT dynamics of complex molecular  systems.\\
 
\begin{acknowledgements}
We acknowledge inspiring comments by D. Ballester and M. Mariantoni. 
P. F.-D. acknowledges funding by the Institute for Quantum Information and Matter, an NSF Physics Frontier Center with support of the Gordon and Betty Moore Foundation, by NSF Grant No. PHY0652914, by the DoD NSSEFF program, by the AFOSR MURI for Quantum Memories, and by Northrop Grumman Aerospace Systems. 
This work is supported by D.P.T (T.R.Prime ministry State Planning Organization) under Project No. 2009K12020 and by National Science Foundation of Turkey under Project No. 109T267 and Project No. 111T285. 
Y. G. gratefully acknowledges support by T\"UB\.{I}TAK Post-Doc Program.
\end{acknowledgements}

%

\end{document}